\documentstyle[aps,prc,epsfig,,twocolumn,multicol]{revtex}

\newlength{\loopepswidth}

\newcommand\LOOPEPSBOX[1]
{
  \begin{minipage}{\loopepswidth}
  \includegraphics[width=\loopepswidth]{#1}
  \end{minipage}
}

\begin{document}

\title{The Coupled Cluster Method
in Hamiltonian Lattice Field Theory: SU(2) Glueballs}

\author{Andreas Wichmann, Dieter Sch\"{u}tte, Bernard C.~Metsch 
        and Vera Wethkamp\thanks{e-mail: {\tt wethkamp@itkp.uni-bonn.de}}\\ 
        {\bigskip}  {\sl Institut f\"ur Theoretische Kernphysik \\
        Nu{\ss}allee 14--16, D--53115 Bonn \\ Germany} \\ {\bigskip} }

\date{\today}

\maketitle

\begin{abstract}
The glueball spectrum within the Hamiltonian formulation of
lattice gauge theory (without fermions) is calculated 
for the gauge group SU(2) and for two spatial dimensions.

The Hilbert space of gauge-invariant functions of the gauge 
field is generated by its parallel-transporters on closed paths 
along the links of the spatial lattice.
The coupled cluster method is used to determine the spectrum of the 
Kogut-Susskind Hamiltonian in a truncated basis. The quality of
the description is studied by computing results from various truncations, 
lattice regularisations and with an improved Hamiltonian. 

We find consistency for the mass ratio predictions within a scaling region
where we obtain good agreement with standard lattice Monte Carlo results. 
 
\end{abstract}

\section{Introduction and survey}
\label{Introduction and survey}

It is the purpose of this paper to present the results of an 
attempt to compute the spectrum of a lattice gauge field theory 
within the Hamiltonian formulation. Our computational framework 
is the coupled cluster method which has been presented and 
discussed in detail in Ref.~\cite{Schutte:1997du}.

We have applied our method to the lattice version of the $SU(2)$ 
Yang-Mills theory in two spatial dimensions which is the simplest 
non-trivial non-abelian lattice gauge field theory.
We consider this model whose dynamics is given by a special 
Kogut-Susskind Hamiltonian as an important test case for  
controlling the coupled cluster method. Although there is no 
experiment to compare with, there exist for this model very reliable 
results for the ground state energy 
(see Ref.~\cite{Hamer:1992ic,Hamer:1996zj}) and for glueball 
masses (see Ref.~\cite{Teper:1999te}) which allow a critical test 
of the coupled cluster predictions.

In the past, several attempts have been made to  compute glueball 
masses within the same model. We mention the light-front appoach 
of van de Sande {\em et al.}~(see Ref.~\cite{Sande}) and the maximal 
tree approach 
of Bishop {\em et al.}~(see Ref.~\cite{Ligterink:2000ug}).

Calculations similar to ours, i.e.~treating the  Kogut-Susskind
Hamiltonian  within the coupled cluster method, were undertaken 
by Llewellyn Smith and Watson (see Ref.~\cite{LlewellynSmith:1993ig}) 
and Guo {\em et al.}~(see Ref.~\cite{Chen:1994ri,Chen:1995ca,Guo:1994vq}).

Within the present work we are able to go far beyond the
limitations of this earlier work:        

\begin{itemize}
\item
When expanding the characteristic correlation functions within a 
localized loop basis (see Ref.~\cite{Schutte:1997du}), we can push 
the expansion up 
to the {\em 7th order}.

\item
We are able to check the validity of the computation of the 
characteristic coupled cluster matrix elements by comparing 
{\em two indepedent formulations} of the local loop space basis 
expansion (see Ref.~\cite{Schutte:1997du,Wichmann}).

\item
We can compare the predictions of {\em different lattice regularizations}
which exist in two spatial dimenions: square, hexagonal and triangular lattice.

\item
We compute glueballs with {\em all possible angular momentum quantum numbers} 
allowed within the lattice formulation, i.e.~$J^P = 0^+,0^-,2^+,2^-,1^\pm$ 
for the square lattice and $J^P = 0^+, 0^-, 3^+, 3^-,1^\pm,2^\pm$ for 
the hexagonal or triangular lattice.

\item
We also test the quality of our results by comparing to those with 
an {\em improved Hamiltonian}.

\end{itemize}

Taking dimensionless {\em mass ratios} we observe a good {\em scaling window}
(around $\beta = 3$ for the non-improved case).
Judging the stability of the results within the  scaling window
by testing the convergence with the order of the coupled cluster 
expansion, by comparing the predictions of the different lattice 
regularisations and the inclusion of improvement, we are able to 
{\em predict} the spectrum of the energetically low-lying glueballs
{\em within certain errors} (see Fig.~\ref{finalresult}).
Within these errors, our prediction of the glueball spectrum 
{\em agrees with that of Teper} (see Ref.~\cite{Teper:1999te}).
Since our results are very encouraging it is, of course, desirable 
to extend these computations to  more physical models.
The $SU(2)$ lattice Yang-Mills theory for $D = 3 + 1$ dimensions has 
been worked out (see Ref.~\cite{Wichmann}), 
but the results are too preliminary 
to be presented. The extension of our model to the $SU(3)$ gauge
group is under way (see Ref.~\cite{Wethkamp}).
The inclusion of quarks, i.e.~the treatment of full QCD, is possible in 
principle, but technically difficult. First attempts using the coupled 
cluster method for such models exist, but only for the Schwinger
model (i.e.~the $1+1$ dimensional $U(1)$ case) (see Ref.~\cite{Fang:2001gq}).

Our paper is organized as follows:

Section \ref{Computational Scheme} defines our computational scheme, 
especially details of the coupled cluster expansion.

Section \ref{The loop space basis}  describes the structure of the 
two local loop space 
expansions (algebraic versus Clebsch-Gordan methods) and how 
the computation of the coupled cluster matrix elements is done.

Section \ref{The Lattice Euclidean Group} gives some details of how 
to project localized wave function 
on good lattice momentum and angular momentum (only lattice momentum zero
is considered within this paper).

Section \ref{Computational details} describes some calculational details, 
especially the truncation 
procedure which is necessary to fix the coupled cluster method.

Section  \ref{Results for the Ground State}
presents results for the {\em vacuum state}. It is shown that within the 
scaling window the vacuum energy agrees with the value  obtained by 
the Greens function Monte Carlo method (see Ref.~\cite{Hamer:1996zj}) 
better than 
within  $0.1\%$ showing that the
coupled cluster method is thus reliable. 

Section \ref{The Glueball Spectrum} gives the results 
for the glueball masses.
The stability of the results within the scaling
window is shown by testing stability with respect to
the order of the coupled cluster expansion 
(Subsection \ref{Results on the square lattice}), 
with respect to improvement (Subsection \ref{Results with improvement}) 
and with respect to the different regularizations 
(Subsection \ref{Results for alternative lattices}).

In the Appendix we discuss the different scales in the Hamiltonian and the
Lagrangian formulation (see Ref.~\cite{Hamer:1996ub}) and study the possibility
to have a {\em scaling region around} $\beta = 3$ for our framework.

\section{Computational Scheme} 
\label{Computational Scheme}

Within the Hamiltonian formulation, the computational framework
of any lattice Yang-Mills theory is given by the Kogut Susskind theory
which has been discussed in detail in Ref.~\cite{Schutte:1997du}. 
In summary, the structure is the following: 

The Kogut Susskind wave functions {} $\Psi(U) = \Psi(U_1, \ldots, U_N)$ 
depend on link variables $U_l$ ($l = 1, \ldots, N$) which are elements 
of the gauge group $SU(N_c)$. $N$ is the number of oriented links in 
a  lattice (with $d$ spatial dimensions).
$N$ is related to the finite volume chosen. 
Our many-body technique allows us to perform the final calculations in
the infinite volume limit $N \rightarrow \infty$.
Our concrete calculations are done for the case $N_c = 2$ and $d = 2$.

The idea of the coupled cluster method is to reformulate the 
eigenvalue problem $H \Psi = E \Psi$ as equations for the ground 
state correlation function $S$ and for the excitation operators $F$,
following from the ansatzes $\Psi_0(U) = \exp{S(U)}$ for the ground state 
and $ \Psi(U) = F(U) \exp{S(U)}$ for excited states.

If $H = 1/(2a)(\sum_{lc}g^2 E^2_{lc} -2/g^2V)$ is the Kogut Susskind 
Hamiltonian (we use the notations $(lc)$ for link-colour quantum numbers,
$E_{lc}$ for  the  momentum operators conjugate to $U_l$ and $V$ for 
the plaquette term) this yields the non-linear equation

\begin{eqnarray}
\sum_{lc} (S_{{lc}{lc}} + S_{lc} S_{lc} ) - \frac{2}{g^4} V 
       &=& \frac{2a}{g^2} E_0
\label{e1}
\end{eqnarray}

for $S$ and  the linear equation

\begin{eqnarray}
\sum_{lc} (F_{{lc}{lc}} + 2 S_{lc} F_{lc}) 
     &=& \frac{2a}{g^2} (E-E_0) F
\label{e2}
\end{eqnarray}

for the excitation operators $F$. We use the notation that 
$f_{lc}$ abbreviates ``link variable derivatives''
of  functions $f(U)$:

\begin{eqnarray*}
f_{lc} &=& \left[E_{lc},f\right]\\
f_{{lc}{lc}} &=& \left[E_{lc},\left[E_{lc},f\right]\right].
\end{eqnarray*}
 
This (rigorous) form of the eigenvalue problem manifestly guarantees
the correct volume dependencies
of both the ground state energy $E_0$ 
and the excitation energies $E - E_0$ (see Ref.~\cite{Schutte:1997du}).

Within this paper we will consider excitations corresponding to glueballs
with arbitrary lattice angular momentum and parity,
but with momentum zero. Thus the eigenvalues are interpreted
as glueball masses. We may then write  

\begin{eqnarray*}
F& = &   F^{k,\nu}_\sigma = \sum_\nu {\tilde \Pi}^k_{\nu\sigma}  
                                  F^{k,\nu}_{\rm int}\\
S &=& {\tilde \Pi}^0 S_{\rm int}
\end{eqnarray*}

where $\tilde \Pi^k_{\sigma\sigma}$ is the projection operator on states
with vanishing (lattice) momentum and  angular momentum-parity 
$k$ (e.g.~$k = J^P  =  0^+,0^-,2^+,2^-,1^\pm$ for the square lattice).
$\nu,\sigma$ are magnetic quantum numbers, for more details see Section 
\ref{The Lattice Euclidean Group}. 
We abbreviate ${\tilde \Pi}^{0^+}_{00} = {\tilde \Pi}^0$.

The essential point of the coupled cluster method is
that it can be shown that the functions $S_{\rm int}$ 
and $F_{\rm int}$ are {\em localized} (see Ref.~\cite{Schutte:1997du})
so that it is plausible to {\em expand} these intrinsic functions  
with respect to a localized, gauge invariant, linked, standard  
basis $\chi^\alpha$ in the form

\begin{eqnarray*}
S_{\rm int}(U) &=& \sum_\alpha S_\alpha \chi^\alpha(U)\\
F^{k,\nu}_{\rm int}(U) &=& \sum_\alpha F^{k,\nu}_\alpha \chi^\alpha(U)
\end{eqnarray*}
 
and to define  the approximation to the eigenvalue problem by 
a {\em truncation} of this basis.

Introducing the constant function via $\chi^0 = 1$, the ``plaquette
function''  by putting $\tilde \Pi_0 \chi^1 = 4 V$ and using the strong coupling 
structure $ \sum_{lc}  \chi^\alpha_{lc lc} = \epsilon_{\alpha} \chi^\alpha$ 
which is true for any convenient basis, the coupled cluster equations
(\ref{e1}) and (\ref{e2})  become equations for the coefficients 
$S_\alpha$ and $F^{k,\nu}_\alpha$: 

\begin{eqnarray}
\epsilon_\alpha S_\alpha + \sum_{\beta,\gamma}  
      C^{\beta,\gamma,0}_{\alpha,0}(k = 0^+) &S_\gamma &S_\beta\nonumber \\
&=& \frac{1}{2 g^4} \delta_{\alpha 1} 
           +\frac{aE_0}{4 N g^2} \delta_{\alpha 0}\label{e3}\\
\epsilon_\alpha F^{k,\nu}_\alpha + 2 \sum_{\beta\gamma,\nu'} 
      C^{\beta\gamma,\nu}_{\alpha,\nu'}(k) &S_\gamma &F^{k,\nu'}_\beta\nonumber\\
&=& {2a \over g^2}(E-E_0) F^{k,\nu}_\alpha \label{e4}
\end{eqnarray}

Here, the coupled cluster matrix elements 
$C^{\gamma\beta,\nu}_{\alpha,\nu'}(k)$
are given by the following prescription:

Define first numbers $c^{\beta\gamma}_{\alpha uw}$
which are related to the action $T(u)$ ($u \in G_E$) of the
lattice Euclidean group $G_E$ (see Section 
\ref{The Lattice Euclidean Group} for details)
on the basis $\chi^\alpha$ and its products by

\begin{eqnarray}
\chi^\beta T(u) \chi^\gamma
      &=& \sum_{\alpha,w\in G_E} 
         c^{\beta\gamma}_{\alpha uw} T(w)\chi^\alpha \label{e5}
\end{eqnarray}

Here only those cases have to be considered where the 
functions $\chi^\beta$ and $T(u) \chi^\gamma$ have a common 
link variable, i.e.~only a finite number of possible values for the
Euclidean group elements $u,w$ occurs in Eq.~(\ref{e5}).

The coupled cluster matrix elements are then related to the matrix 
elements $D^k_{\nu\nu'}(b)$ of the representations 
of the lattice rotations $b$ 
($u = (b,{\bf a})$, where ${\bf a} =$ lattice translation,
see Section \ref{The Lattice Euclidean Group}) by

\begin{eqnarray}
C^{\gamma\beta,\nu}_{\alpha,\nu'}(k)
&=& \frac{1}{2}(\epsilon_{\alpha} - \epsilon_{\beta} - \epsilon_{\gamma})
\sum_{u=(b,{\bf a}),\atop w=(b',{\bf a'})} 
  D^k_{\nu,\nu'}(b^{-1} b') c^{\gamma\beta}_{\alpha uw}\label{e6} 
\end{eqnarray}

\section{The loop space basis} 
\label{The loop space basis}       

The basis $\chi^\alpha$ which we use for 
our calculations 
is systematically generated from Eq.~(\ref{e5})
in two steps.

In a first step, certain subspaces $H^{\delta,k}$ of dimension 
$m_{\delta,k}$ are generated from geometrically inequivalent 
plaquette products. The second step then is the construction and
the handling of orthogonal basis systems of these
subspaces, especially the computation of the
coupled cluster matrix elements. This is done in two alternative 
ways, by algebraic and by Clebsch Gordan techniques.

The subspace $H^{\delta,k}$ is {\em generated} from a
function $\Lambda_G^{\delta,k}$ which is given by a linked,
standard $\delta$-fold {} plaquette product of the type

\begin{eqnarray}
\Lambda_G^{\delta,k}&=&\chi^1 T(u_2(\delta,k))
\chi^1 \ldots T(u_\delta(\delta,k))\chi^1\label{e7}\\
k&=&1, \ldots, n_\delta\nonumber
\end{eqnarray}

where $u_\lambda(\delta,k)$ ($\lambda=2,\ldots,\delta$) are suitable 
elements of the lattice 
Eucildean group. For $\delta = 0,1$ we define  $n_0 = n_1 = 1$, 
$\Lambda_G^{0,1} = \chi^0$, $\Lambda_G^{1,1} = \chi^1$.

The ``{\em order}'' of the functions in the subspace  $H^{\delta,k}$
is defined to be $\delta$. In general, the quantum number $k$ 
characterizes the different, geometrically independent
possibilities for the construction of the plaquette products
of the same order whose total number is $n_\delta$.

Up to 7th order we have for our two-dimensional $SU(2)$ case 

$n_\delta = 1, 1, 2, 4, 12, 35, 129, 495 $ 
for the square lattice,

$n_\delta = 1, 1, 2, 5, 15, 53, 235, 1125$ 
for the hexagonal lattice and

$n_\delta = 1, 1, 2, 3, 8, 17, 54, 162$
for the triangular lattice, for $\delta=0,\ldots,7$.

The functions $\Lambda_G^{\delta, k}$ are characterized by simple 
loop patterns exemplified up to 4th order in Tabels 
\ref{Tab: EmpireSquareLattice}, \ref{Tab: EmpireHexagonalLattice} 
and \ref{Tab: EmpireTriangularLattice}.
Here, also the dimensions 
$ m_{\delta,k}$ of the corresponding subspaces $H^{\delta,k}$ are given.

Each function $\Lambda_G^{\delta,k}$ ($\delta, k$ fixed) generates 
an  orthogonal basis $\chi^{(\delta, k, r)}$, $r = 1,..,m_{\delta,k}$,
of $H^{\delta,k}$, and taking all quantum numbers $(\delta,k,r)$ and 
the limit $\delta \rightarrow \infty$, this basis becomes complete for
the expansion of localized functions.

There are two ways to define the basis of $H^{\delta,k}$
explicitely.

\subsection{The algebraic method} 

The idea of this procedure is to act on  $\Lambda_G^{\delta,k}$ 
with a suitable set of commuting Casimir operators of a lattice gauge group.
Hereby, this lattice gauge group $SU(N_c))^M$ is finite dimensional, 
where $M(\delta,k)$ is the number of link variables contained 
in $\Lambda_G^{\delta,k}$.

This yields a generating system for the space $H^{\delta,k}$
and diagonalizing the corresponding Casimir matrices
defines the orthogonal basis. The relation to the generating system
allows  then to compute also the coefficients
$c^{\beta\gamma}_{\alpha uw}$ (see Ref.~\cite{Schutte:1997du}
for details and examples).

The drawback of this method to construct  the basis 
$\chi^\alpha = \chi^{(\delta, k, r)}$ is that one has to take 
care of possible linear dependencies between the generated loop 
space functions. In previous investigations 
(see Ref.~\cite{LlewellynSmith:1993ig,Chen:1994ri,Chen:1995ca,Guo:1994vq,Schutte:1997du})
this problem was solved by exploiting the
Cayley Hamilton relationship between matrices.
We have used in this connection a much simpler procedure: 
If a set  $f_1(U), \ldots, f_n(U)$ contains only $m$ ($m \leq n$) independent
functions, the matrix $f_i(U^k)$ ($i,k = 1, \ldots, n$) has for suitable 
{\em fixed} variables $(U^1, \ldots, U^n)$ exactly the rank $m$.
Our experience is that statistically chosen variable sets 
$(U^1, \ldots, U^n) \in SU(2)$ are suitable in this sense. 
The linear relation between the functions $f_i$ is then easily constructed 
and the dependent functions can be eliminated.

This works quite well up to the 6th order.
For higher orders, however, this method became
too unreliable because of increasing 
numerical errors when checking the linear
dependencies in this way.

\subsection{The Clebsch-Gordan method}

Each basis function $\chi^\alpha$ generated by the
algebraic  method can be characterized by a set of angular momentum 
quantum numbers  adjoined to a certain plaquette pattern  
related to the corresponding function $\Lambda_G^{\delta,k}$. 
For the case of a hexagonal lattice, this pattern consists of 
adjoining a certain angular momentum quantum number $j_l$ to each 
link variable occurring in $\Lambda_G^{\delta,k}$, the possible 
values of $j_l$ being given by simple coupling rules. 
For the square (see Ref.~\cite{Schutte:1997du}) or the triangular lattice, 
also ``intermediate''
quantum numbers have to be taken into account.
This can be systemized by replacing the lattice by a certain ``net'' 
containing only three-point vertices and by defining the plaquette 
pattern within this net. We skip here the description of this net,
details can be found in Ref.~\cite{Wichmann}. 

The set ${\{j_l\}}$ of these quantum numbers defines 
$\chi^\alpha = \chi^{\{j_l\}}$ as a well defined function of the 
link variables $U_l$ in terms of Clebsch-Gordan coefficients, 
see Ref.~\cite{Robson:1982ws,Irving:1986hy,Schutte:2001ri,Wichmann}. 
For convenience, the set  $\{j_l\}= (j_1, j_2, \ldots)$ will be assumed 
to be infinite dimensional by putting $j_l =0$ for all links of the net
not occurring  in the plaquette pattern related to $\chi^\alpha $.

This makes it possible to work directly with the orthonormal basis 
$ \chi^{\{j_l\}}$ and to compute the coupled cluster matrix elements 
with the help of standard group theoretical methods.
Generalizing the notion $\{j_l\}$ also to non-standard functions, 
i.e.~putting $ T(u)\chi^{\{j_l\}} =  \chi^{\{j_l'\}}$, the essential 
formula allowing to compute the coefficients 
$c_{\alpha uw}^{\gamma \beta}$ via certain 9$j$-symbols is given by

\begin{eqnarray}
\chi^{\{j_l^1\}}\chi^{\{j_l^2\}} 
& = &\sum_{\{j_l^3\}} c^\chi(\{j_l^1\},\{j_l^2\},\{j_l^3\}) \chi^{\{j_l^3\}},
\label{e8}
\end{eqnarray}

\begin{eqnarray}
c^\chi(\{j_l^1\},\{j_l^2\},\{j_l^3\}) 
&=&
\prod_{{\mbox{\small positively}}\atop {\mbox{\small oriented \em l}}} 
      \hat j_l^1 \hat j_l^2 \hat j_l^3 
      \nonumber\\
&&\times \prod_{{\mbox{\small \em v}}}\left\{
   \begin{array}[c]{lll}
   j^1_{{\small l}(v,1)} & j^2_{{\small l}(v,1)} & j^3_{{\small l}(v,1)}\\
   j^1_{{\small l}(v,2)} & j^2_{{\small l}(v,2)} & j^3_{{\small l}(v,2)}\\
j^1_{{\small l}(v,3)} & j^2_{{\small l}(v,3)} & j^3_{{\small l}(v,3)}
\end{array} \right\},
\label{e9}
\end{eqnarray}

where $\hat j =\sqrt{2j+1}$.

In Eq.~(\ref{e8}) the (in each case finite) sum runs only over those
sets ${\{j_l^3\}}$ which obey for each $l$ the selection rule 
$|j_l^1 - j_l^2| \leq j_l^3 \leq j_l^1 + j_l^2$ and the condition 
$j_l^1 \neq 0$ or $j_l^2 \neq 0$.

In Eq.~(\ref{e9}), we have used the notation 
$\{l(v,1), l(v,2), l(v,3)\}$ 
for the three links $l$ adjacent to the vertex $v$.

For the application of the coupled cluster equation
{\em including truncation} one has to know the
{\em order} of the function $\chi^{\{j_l\}}$,
i.e.~to which subspace $H^{\delta,k}$ it belongs - the plaquette pattern
of $ {\{j_l\}}$ does not give this 
uniquelly. This order is, however, easily obtained
when constructing the basis $\chi^{\{j_l\}}$ by iterative
Clebsch-Gordan coupling, see Ref.~\cite{Schutte:1997du}
for examples.

\section{The Lattice Euclidean Group}
\label{The Lattice Euclidean Group}        

Any explicit calculation involves the treatmant of the
lattice Euclidean group $G_E$ which consist of
translations and rotatons, $G_E = G_T \otimes_s G_R$,
leaving the (infinite) lattice invariant.

\subsection{Lattice Translations}
 
Translations ${\bf x} \to {\bf x} +{ \bf a}$
are given by translation vectors 

\begin{eqnarray*}
{\bf a} &=& a_1 {\bf e}_1 + a_2  {\bf e}_2
\nonumber
\end{eqnarray*}

where 
${\bf e}_1 = \left( \begin{array} {c}
                    1 \\
                    0 \\
             \end{array} \right)$.
For the other unit vector we have that
${\bf e}_2 = \left( \begin{array} {cc}
                    0 \\
                    1 \\
              \end{array} \right)$
for a square lattice and
${\bf e}_2 = \left( \begin{array} {cc}
                   \cos(\frac{\pi}{3}) \\
                   \sin(\frac{\pi}{3}) \\
              \end{array} \right)$
for the hexagonal or triangular lattice, respectively.

All distances will be measured in units of the lattice spacing.
The numbers $a_1,a_2$ are then integers which are arbitrary
for the square or the triangular lattice and obey
a simple restriction for the hexagonal lattice (see Ref.~\cite{Wichmann}).

\subsection{Lattice Rotations}

General Euclidean transformation are given by a pair
$(b,{\bf a})$  mapping ${\bf x} \to b {\bf x} +{ \bf a}$
where $b \in G_R$ is a lattice rotation.

For the  square lattice, the ``point group'' $G_R$ has
the eight elements $(b_0, ..,b_7)$ which may be written
in terms of the following
$ 2 \times 2 $ matrices

\begin{eqnarray*} 
b_n &=&\left( \begin{array}[c]{ll}
               \cos (n\frac{\pi}{4}) & \sin (n\frac{\pi}{4})   \\
             - \sin (n\frac{\pi}{4}) & \cos (n\frac{\pi}{4})
            \end{array} \right)\\
b_{n+4} &=& \left( \begin{array} {cc}
                    0 & 1\\
                    1 & 0\\
             \end{array} \right) b_n \,\, ,n=0, \ldots, 3 \,\,.
\end{eqnarray*} 

For the  {\em hexagonal or triangular lattice} the  group $G_R$ has
the 12 elements $(b_0, ..,b_{11})$ which we write in the form
\begin{eqnarray*} 
b_n &=&\left( \begin{array} {cc}
             \cos (n\frac{\pi}{3}) & \sin (n\frac{\pi}{3})   \\
           - \sin (n\frac{\pi}{3}) & \cos (n\frac{\pi}{3})
        \end{array} \right)\\
b_{n+6} &=& \left( \begin{array} {cc}
                     1 & 0\\
                     0 & -1\\
            \end{array} \right) b_n \,\, , n=0, \ldots, 5\,\,.
\end{eqnarray*} 

\subsection{Irreducible Representations}

We now list the irreducible representaions (irreps) of $G_R$ 
(see e.g.~Ref.~\cite{Altmann}).

For the  square lattice ({\em hexagonal or triangular lattice}) 
we have 5 ({\em 6}) inequivalent irreducible representations (irreps) 
$D^k$, 4 one-dimensional irreps labelled $A_1,A_2, B_1, B_2$ and 
one ({\em two}) two-dimensional irreps labelled E (${\em E}_1,{\em E}_2$).
In the continuum limit, these irreps will correspond to the lowest
spins $0^+,0^-,2^+,2^-$ and $1^\pm$ 
(${\em 0}^+,{\em 0}^-,{\em 3}^+,{\em 3}^-, {\em 1}^\pm$ and ${\em 2}^\pm$), 
respectively. Note that, in the continuum limit, the eigenvalues
for the quantum numbers $J^+$ and $J^-$ have to be degenerate for $J \neq 0$
for a rotational invariant Hamitonian. Explicitly, the irreps read for 
the one-dimensional cases ($n=0, \ldots, 7)$ for the square lattice
({\em n=0, \ldots, 11 for the hexagonal or triangular lattice})

$D^{A_1}(b_n) = 1$ 

$D^{A_2}(b_n) = \det(b_n)$ 

$D^{B_1}(b_n) = (-1)^n \det(b_n)$ 

$D^{B_2}(b_n) = (-1)^n$

The two-dimensional irreps  read for the square case

$D^{E}(b_n) = b_n$ $(n = 0, \ldots, 7)$,

and for the {\em hexagonal or triangular case}

$D^{E_1}(b_n) = b_n$ 

$D^{E_2}(b_n) = (-1)^n b_n$ $(n = 0, \ldots, 11)$

\subsection{Projections on  Lattice Momenta and Angular Momenta}

The group $G_E$ acts on the functions $\chi^{\{j_l\}}$
via

\begin{eqnarray*}
T(u) \chi^{\{j_l\}} &=& \chi^{\{j_{l'}\}}\\
              l'&=&u^{-1}l
\end{eqnarray*}

for $u\in G_E$.

This allows to define  projection operators
on good momentum ${\bf p}$ within the Brillouin zone
by

\begin{eqnarray*}
\Pi_{\bf p}^T &=& \sum_{\bf a} e^{i {\bf pa}} T(0,{\bf a}).
\end{eqnarray*}

States with fixed lattice angular momentum may be 
constructed by the operators

\begin{eqnarray}
\Pi^k_{\nu \sigma} &=& N_k \sum_{n} D^k_{\nu \sigma}(b_n^{-1}) T(b_n,0),
\label{e10}
\end{eqnarray}

where the normalization factor $N_k = \dim D^k/|G_R|$
guarantees the projector property

\begin{eqnarray*}
\Pi^k_{\nu \nu} \Pi^{k}_{\nu \nu} &=&
 \Pi^k_{\nu \nu}.
\end{eqnarray*}

The index $\sigma$ in Eq.~(\ref{e10}) corresponds to
a ``magnetic'' quantum number since we have 

\begin{eqnarray}
T(b)\Pi^k_{\nu \sigma}
&=& \sum_{\sigma'} D^k_{\sigma,\sigma'}(b) \Pi^k_{\nu \sigma'}.
\label{e11}
\end{eqnarray}

Thus for the eigenvalue problem of a rotationally invariant
Hamiltonian $\sigma$ can be fixed, e.g.~to  $\sigma = 0$.

Eq.~(\ref{e11}) also shows that the operator $\Pi^k_{\nu \sigma}$
generates the representation $D^k$ twice if it is
two-dimensional. Thus the index $\nu$ is not a
``good'' quantum number, the Hamiltonian will in
general mix states will different values of $\nu$.

For our concrete calculations we restricted
the computation to eigenstates
with momentum zero.  This simplifies the
work because $\Pi^T_0$ commutes with the operators
$\Pi^k_{\nu \sigma}$.  A basis for the expansion of the  
correlation functions $S$ or $F$ is then generated from
the localized basis $\chi^\alpha$ via

\begin{eqnarray}
\{ {\tilde \Pi}^k_{\nu 0} \chi^\alpha &=& 
\Pi^k_{\nu 0} \Pi^T_0 \chi^\alpha | \alpha \in N, 
0 \leq \nu \leq \dim D^k \}\,\,.
\label{e12}
\end{eqnarray}

The states of Eq.~(\ref{e12}) form an orthogonal
basis if the functions $\chi^\alpha$ 
are orthogonal {\em and} if - for the given $k$ - states are
left out which are projected to zero under
$ {\tilde \Pi}^k_{\nu 0}$. This may occur if the
loop space pattern related to $\chi^\alpha$ is
too symmetric, i.e.~if this state is invariant
under a subgroup of the lattice rotation group.
Tables \ref{Tab: PiChiSquareLattice}, \ref{Tab: PiChiHexagonalLattice} 
and \ref{Tab: PiChiTriangularLattice} give a survey on the 
remaining dimension
of the relevant subspaces up to the 7th order. Note that for higher
angular momenta only calculations in orders with at least $\delta\ge5$ 
make sense.

\section{Computational details} 
\label{Computational details}    

With the definition of Eq.~(\ref{e12}) for the basis expansion
of the correlation functions $S$ and $F$,
it is straightforward to show the validity of
the form  Eq.~(\ref{e6}) of the coupled cluster
matrix elements.

For the numerical study, we have set
up a computer program which organized
efficiently the handling of the loop space
and of the lattice Euclidean group.
For the computation of the coupled
cluster matrix element, we used
up to 6th order both the algebraic and the
Clebsch Gordan method. We found 
a perfect agreement which was 
a comforting assurance of the
validity of our computational methods.
Details are described in Ref.~\cite{Wichmann}.

For the concrete calculations,
we still have to define a truncation prescription.
For the results presented below,
we follow the proposal of 
Guo {\em et al.}~(see Ref.~\cite{Chen:1994ri,Chen:1995ca,Guo:1994vq})
which yields actually  the simplest
set of equations. In this case one puts
in the order $\delta$

\begin{eqnarray}
c^{\alpha_1,\alpha_2}_{\alpha_3,uw} = 0
\;\;&\mbox{ for }&\;\;\; \delta(\chi^{\alpha_1}) + \delta(\chi^{\alpha_2}) > \delta.
\label{Truncation}
\end{eqnarray}

Since the expansion of the function $S$ starts with
$\delta = 1$, our calculations go up to 8th order
if the expansion of the correlation functions
is pushed to 7th order, as we do within this work.

We have also studied alternative truncations (see Ref.~\cite{Wichmann}),
but we could not find any prescription which was better
than Eq.~(\ref{Truncation}).

\section{Results for the Ground State}
\label{Results for the Ground State}

It is a special feature of the Hamiltonian 
formulation  that it provides both the energy and the wave
function of the vacuum state which
is the ground state of the  Hamiltonian.

Standard lattice Monte Carlo calculations
do not give results for the vacuum energy
density. There exist, however, computations
within the strong coupling expansion (see Ref.~\cite{Hamer:1992ic}) and
very reliable Green's function Monte Carlo results 
(see Ref.~\cite{Hamer:1996zj}).

\begin{figure}
\input{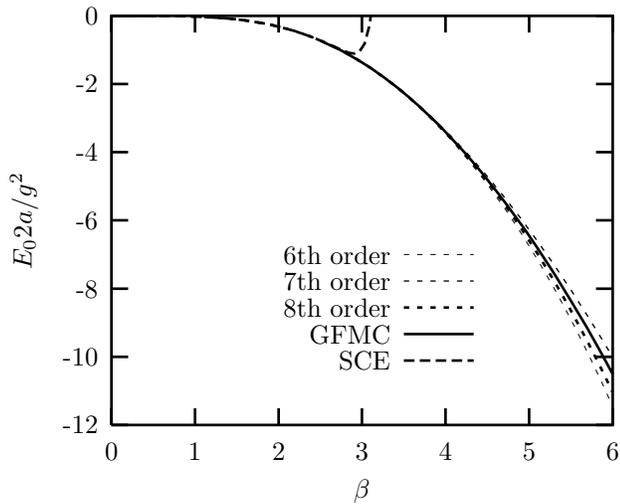}
\caption{Ground state energy density as a function 
       of the inverse
       coupling $\beta=\frac{4}{g^2}$ up to 8th order. For comparison:
       Strong coupling expansion SCE~\protect \cite{Hamer:1992ic} 
      and Green's function Monte-Carlo GFMC method~\protect \cite{Hamer:1996zj}.}
\label{Fig:Ground}
\end{figure}

In Figure \ref{Fig:Ground} we compare our coupled cluster
results up to 8th order to the results of the
other methods. As coupling variable we use the standard expression
$\beta = 4/g^2$.
We see that we obtain excellent results (better than 0.1\%)
in our scaling region ($\beta \approx 3$, see below),   
we have a good quality and
convergence up to $\beta \approx 4$,
but our method breaks down for large $\beta$
where the Green's function Monte Carlo is 
still valid.

An important feature of our results is that the
validity of the coupled cluster method clearly 
goes beyond the range of the strong coupling
expansion which breaks down at $\beta \approx 3$.
(Our figure gives the result of the 18th order of strong
coupling perturbation theory!) We consider
this improvement as a necessary condition for obtaining
continuum limit physics.

We should stress  that the determination
of the vacuum operator $S(U)$ - by solving Eq.~(\ref{e3})  iteratively -
turns out to be much simpler and faster than that of
the excitation operator $F(U)$.
Thus we do not see any special
difficulty with the fact that the
Hamiltonian formulation also involves
the determination of the vacuum state.
The coupled cluster formulation deals
with this problem apparently quite effectively.

In this sense our framework seems to be
quite orthogonal to the light front formulation
which is based on the assumption that
the simplicity of the vacuum could be
helpful for the determination of the spectrum.

\section{The Glueball Spectrum} 
\label{The Glueball Spectrum}       

The eigenvalues of the solutions of Eq.~(\ref{e4}) 
have the interpretion of  (approximate)  glueball masses
with angular-momentum $J^P = k$.

\subsection{Results on the square lattice}
\label{Results on the square lattice}

\begin{figure}
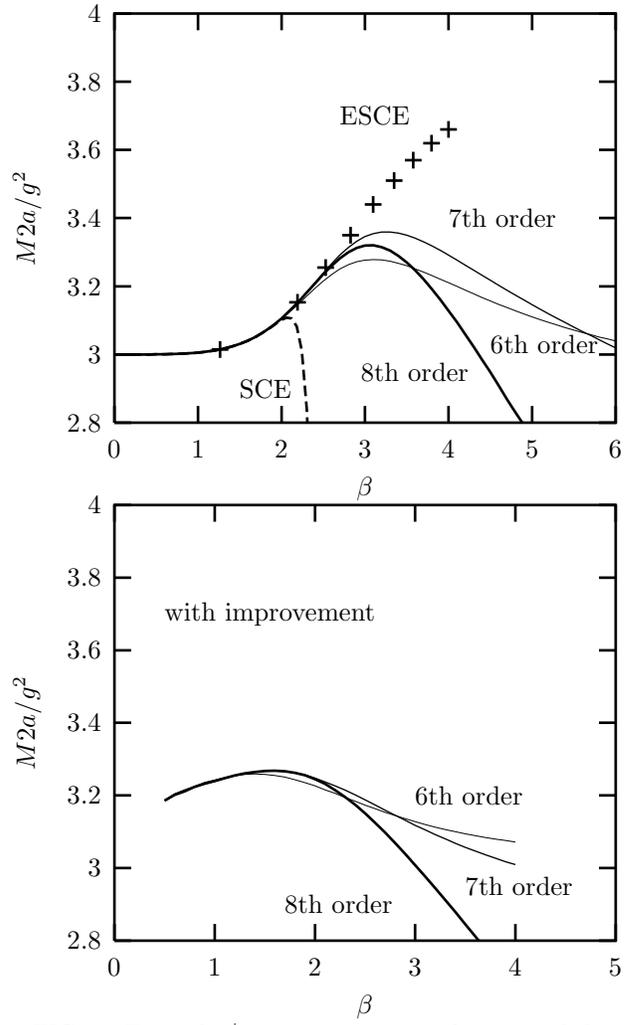

\begin{minipage}[]{8.5cm}
\begin{minipage}[t]{8.5cm}
    \input{0plusglueball.latex}
\end{minipage}
\begin{minipage}[t]{8.5cm}
  \input{0plusglueballimproved.latex}
\end{minipage}
\caption{Excited $0^+$ state energy 
    as a function of the inverse coupling $\beta=\frac{4}{g^2}$ 
    up to 8th order without and with improvement. 
    For comparison: Strong coupling expansion SCE 
    and Extrapolation of the strong coupling 
    series ESCE with Shafer and integrated differential 
    approximants~\protect \cite{Hamer:1992ic}.}
\label{Fig:0plusglueball}
\end{minipage}
\end{figure}

An example of the convergence of the lowest $0^+$ glueball
mass up to the 8th order of the coupled cluster
expansion is given in Fig.~\ref{Fig:0plusglueball} (upper figure). We see that there
is perfect agreement with Hamer's results of 
the strong coupling expansion (see Ref.~\cite{Hamer:1992ic}) 
up to $\beta \approx 2$
which breaks down for larger $\beta$. Also
Hamer's ESCE method (see Ref.~\cite{Hamer:1992ic}) becomes apparently
unreliable for $\beta \geq  3$.

Obviously, we are able to estimate in the region $\beta \approx 3$
the value of the glueball mass within errors.

The problem is whether this region of couplings contains
already continuum physics. This we judge from the
occurence of an approximate scaling window for
mass ratios.

\begin{figure}
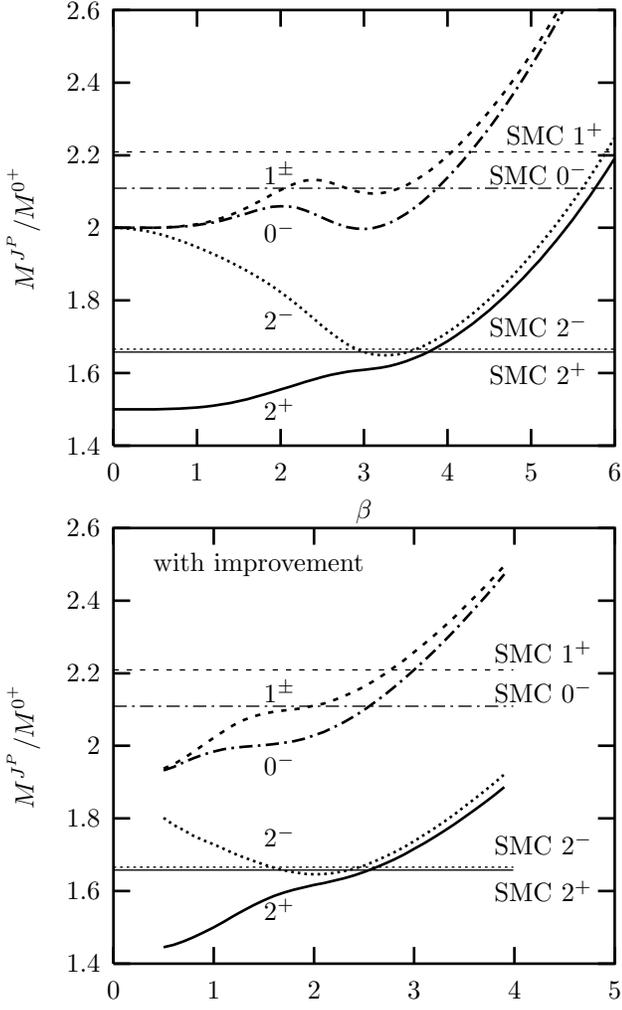

\begin{minipage}[]{8.5cm}
\begin{minipage}[t]{8.5cm}
     \input{massratios.latex}
\end{minipage}
\begin{minipage}[t]{8.5cm}
\input{massratiosimproved.latex}
\end{minipage}
\caption{Mass ratios of the $J^P= 0^{-}$, $J^P=2^+$ , $J^P=2^-$ 
       and the $J^P=1^{\pm}$ 
      state relative to the $0^+$ state as a function of the inverse coupling 
       $\beta=\frac{4}{g^2}$ in 8th order without and with improvement. 
       For comparison: Standard Monte 
       Carlo SMC results~\protect \cite{Teper:1999te} for $\xi=1$
       (only scaling window result). $\xi$ is the anisotropic factor 
        $\xi=\frac{a_s}{a_t}$, $a_{s(t)}=$ lattice spacing in space 
       (time) direction.}
\label{Fig:massratios}
\end{minipage}
\end{figure}

\begin{figure}
\input{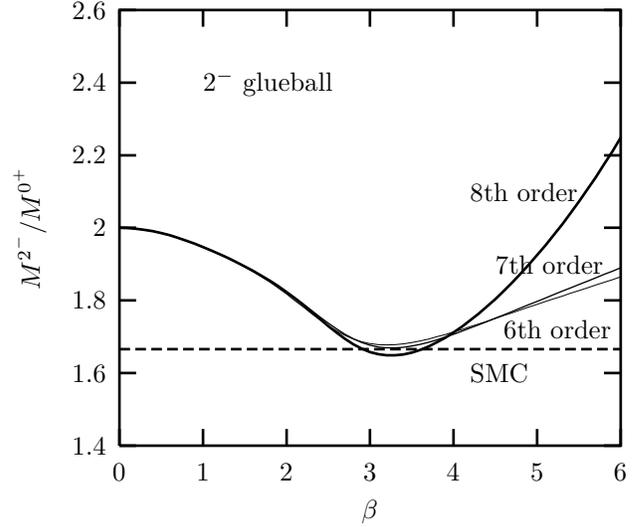}
\caption{Mass ratios of the $J^P=2^-$ state relative to 
     the $0^+$ state as a function of the inverse coupling 
      $\beta=\frac{4}{g^2}$ 
     in 6th, 7th and 8th order to show the convergence with the order.
      For comparison: Standard Monte Carlo SMC 
     results~\protect \cite{Teper:1999te} 
     for $\xi=1$ (only scaling window result). $\xi$ is the anisotropic 
       factor 
        $\xi=\frac{a_s}{a_t}$, $a_{s(t)}=$ lattice spacing in space 
       (time) direction.}
\label{Fig:massratios2minus}
\end{figure}

\begin{figure}
\input{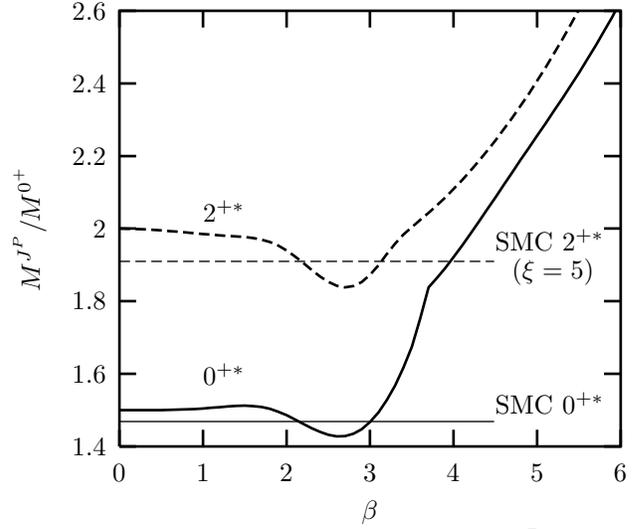}
\caption{Mass ratios of the first excitations $J^P= 0^{+*}$ 
         and $J^P= 2^{+*}$ glueball states in 8th order.
           For comparison: Standard Monte Carlo SMC
         results~\protect \cite{Teper:1999te} for 
         $\xi=1$ (only scaling window result). 
       $\xi$ is the anisotropic factor 
        $\xi=\frac{a_s}{a_t}$, $a_{s(t)}=$ lattice spacing in space 
       (time) direction.}
\label{Fig:massratiosexcited}
\end{figure}

In Fig.~\ref{Fig:massratios} (upper figure) we present mass ratios of the 
different lowest $J^P$ states of the square lattice relative to the mass 
of the $0^+$ state in $8$th order. As an example of the convergence with 
the order we give in Fig.~\ref{Fig:massratios2minus} for the three highest 
orders the mass ratios of the $2^-$ state versus the $0^+$ state.  
We compare our data with the scaling window results of Teper 
(see Ref.~\cite{Teper:1999te}). As an example of mass ratios of higher 
excitations we show in Fig.~\ref{Fig:massratiosexcited} 
that of the $0^{+*}$ and the $2^{+*}$ state, because here a comparison 
with Teper is also possible.

We assume the scaling window to be at $\beta \approx 3$. This allows us 
to estimate the glueball spectrum within certain errors
by comparing the prediction of the
7th and 8th order at $\beta=3$ , see Table \ref{Tab: MtoM0plus}.

Also included are the predictions of the improved Hamiltonian 
(see Subsection \ref{Results with improvement}) and of alternative lattices 
(see Subsection \ref{Results for alternative lattices}).

\begin{figure}
\input{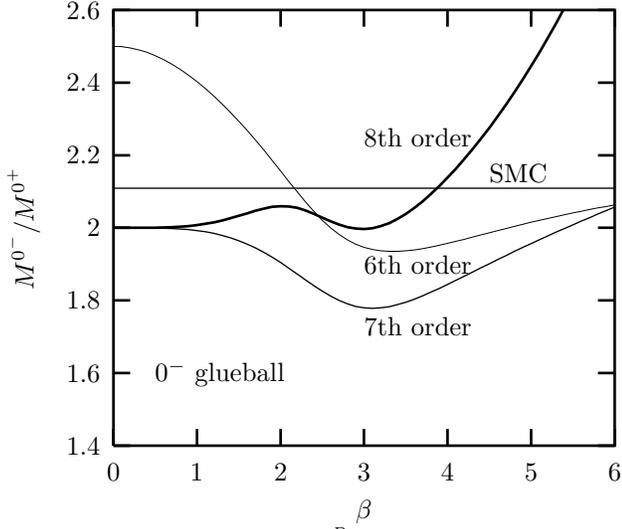}
\caption{Mass ratios of the $J^P=0^-$ state relative to 
     the $0^+$ state as a function of the inverse 
     coupling $\beta=\frac{4}{g^2}$ 
     in 6th, 7th and 8th order. 
     This sector is dominated by $2$-cluster states. 
     For comparison: Standard Monte Carlo SMC
      results~\protect \cite{Teper:1999te} for $\xi=1$ 
        (only scaling window result). $\xi$ is the anisotropic factor 
        $\xi=\frac{a_s}{a_t}$, $a_{s(t)}=$ lattice spacing in space 
       (time) direction.}
\label{Fig:massratios0minus}
\end{figure}

For the $0^-$ sector  we have 
to remark that because the glueball in question lies in the continuum of two 
(lowest) $0^+$ glueball states the results are not very reliable.
This can be seen having a look on the wave functions of this states.
In the $0^-$ sector extending the calculations from $6$th to $7$th and $8$th
order a new state occurs (see Fig.~\ref{Fig:massratios0minus}).
In the strong coupling limit ($\beta=0$) it has an eigenvalue of twice the
eigenvalue of the $0^+$ glueball and its wave function is exactly the product
of two glueball excitation operators: the glueball excitation $F$ is given 
by the product of two plaquettes ({\it two-cluster state}):

\begin{eqnarray*}
F_{0^+} (\beta=0)=&&
\LOOPEPSBOX{looppicture-glueballanalysis-geo0-degas1.eps}, 
7\textrm{th order}\\
F_{0^-} (\beta=0)=&&
\LOOPEPSBOX{looppicture-glueballanalysis-geo0-degas1114.eps}, 
7\textrm{th order}\,\,.
\end{eqnarray*}

As is the strong coupling limit, the scaling region ($\beta=3$) is 
dominated by {\it two-cluster states}:

\begin{eqnarray*}
F_{0^{-}}(\beta=3.0):
&&\begin{array}{c}
\LOOPEPSBOX{looppicture-glueballanalysis-geo0-degas12770.eps}\\
20.25\%
\end{array}
+\begin{array}{c}
\LOOPEPSBOX{looppicture-glueballanalysis-geo0-degas12641.eps}\\
10.67\%
\end{array}
+\begin{array}{c}
\LOOPEPSBOX{looppicture-glueballanalysis-geo0-degas1114.eps}\\
10.57\%
\end{array}
+\begin{array}{c}
\LOOPEPSBOX{looppicture-glueballanalysis-geo0-degas342.eps}\\
5.86\%
\end{array}\\
&&+\begin{array}{c}
\LOOPEPSBOX{looppicture-glueballanalysis-geo0-degas137.eps}\\
4.11\%
\end{array}
+\begin{array}{c}
\LOOPEPSBOX{looppicture-glueballanalysis-geo0-degas61.eps}\\
4.02\%
\end{array}
+\begin{array}{c}
rest\\
44.52\%
\end{array}, 8\textrm{th order}\,\,.\\
\end{eqnarray*}

This feature of the wave functions appears in the whole $0^-$ sector and also
in the $1^{\pm}$ sector for the $1^{\pm*}$ state.

The quality of the result on the square lattice is now tested 
by comparing to other computational schemes
yielding equivalent descriptions in the continuum limit.

\subsection{Results with improvement}
\label{Results with improvement}

The definition of an improved action (see Ref.~\cite{Lepage:1998id})  
or an improved
Hamitonian (see Ref.~\cite{Luo:1999dx}) is not unique. Within our work,
we have studied the simplest choice, namely a tadpole
improvement of the plaquette part of the Hamiltonian.
This is given by the following prescriptions (see Ref.~\cite{Lepage:1998id}).

Replace
\begin{eqnarray*}
  \textrm{tr}(\Box)
  &\quad\to\quad&
  \frac{1}{u_0^4}\left(
  \frac{5}{3} \textrm{tr} ( \Box )
  -
  \frac{1}{6u_0^2} \textrm{tr} (
\setlength{\loopepswidth}{12mm}
\LOOPEPSBOX{looppicture-geo0-picasso5.eps}
  \hspace*{-3.8mm}
  )
  \right)
\end{eqnarray*}

where $u_0$ is related to the vacuum expectation value
of the plaquette via

\begin{eqnarray*}
  u_0^4 &=& \langle \psi_0 | \frac{1}{N_c} \textrm{tr}(\Box) | \psi_0 \rangle 
\end{eqnarray*}

which, in turn, may be computed using the 
Feynman-Hellmann formula

\begin{eqnarray*}
\langle 0 | \textrm{tr}(\Box) | 0 \rangle
&=&\frac{\textrm{d}}{\textrm{d}\lambda} \langle 0 | H 
     + \lambda \textrm{tr}(\Box) | 0 \rangle \big|_{\lambda=0}.
\end{eqnarray*}

Improvement changes the scales of $\beta$. Our results
show that the scaling region is shifted from
$\beta \approx 3$ to $\beta \approx 1.7$,
but the prediction of mass ratios should not change.

Repeating the computation of the ground state with
improvement we find similar convergence properties
as before. The plaquette vacuum expectation value
turns out to be $u_0 \approx 0.84$ for $\beta \approx 2$.

As an example for the quality of the convergence 
with improvement we show in Fig.~\ref{Fig:0plusglueball} (lower figure)
the behaviour 
of the lowest $0^+$ glueball
mass and in Fig.~\ref{Fig:massratios} (lower figure) the structure of the 
scaling window which appear to be improved, indeed.

Taking the predictions of 7th and 8th order
at $\beta =1.7$ we obtain for
the glueball ratios the numbers given in Table \ref{Tab: MtoM0plus}
which show a remarkable agreement to the non-improved
results.

\subsection{Results for alternative lattices} 
\label{Results for alternative lattices}

We have also computed the ground state and the
glueball spectrum for the hexagonal and for the triangular
lattice. Using the ``geometrical'' rescalings

\begin{eqnarray*}
\beta &\rightarrow &\frac{2}{3} \beta\\
E_0 &\rightarrow &\frac{2}{3} E_0
\end{eqnarray*}
we find in both cases perfect agreement of the vacuum energy
to that for the square lattice.

Using the same rescaling for the glueball masses,
we observe again an approximate
scaling region for $\beta \approx 3$.
Since for these new lattices, the scaling
quality for the $2^+$ glueball is better
than that for the $0^+$ case, the numbers presented 
in Table \ref{Tab: MtoM0plus} were estimated of the
mass ratios of the different $J^P$ states relative 
to the $2^+$ glueball and than rescaled to the
$0^+$ state.

For both new lattices, we find 
a new prediction for the $3^+$ and $3^-$ glueball
which turns out to be approximately degenerate in the scaling region,
as it should be.

For the triangular lattice, because of level crossings between
the lowest and the first excited state in the $1^{\pm}$ and in the 
$3^-$ sector we find results which differ from other lattices.

Like on the square lattice also on the hexagonal and the triangular lattice
the $0^-$ sector is dominated by two-cluster states.
In addition the two-cluster states influence also
the first and second excitations in the  $1^{\pm}$ and the $3^{\pm}$ 
sector on the hexagonal lattice.

\begin{figure}
\input{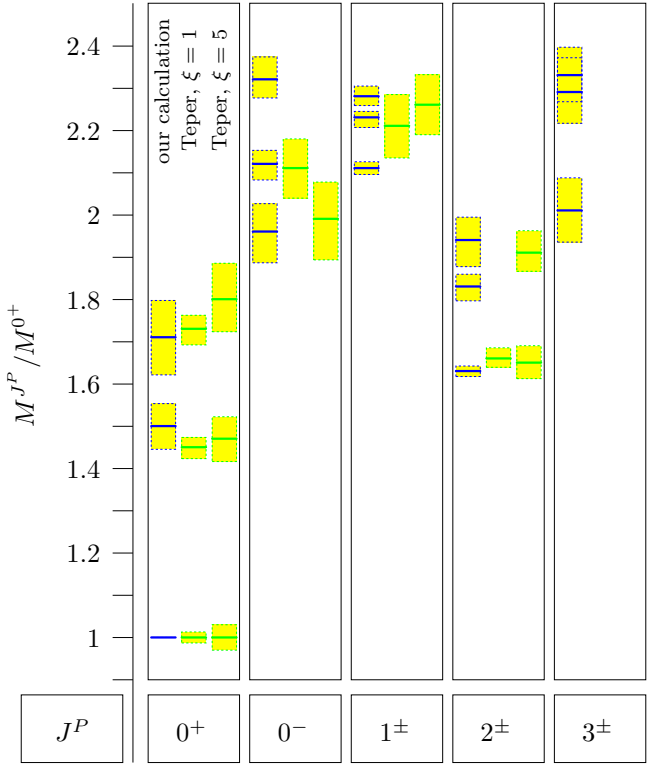}
\caption{Final estimate of the glueball spectrum in units of the $0^+$ 
         glueball. The bars in the left column are mean values 
         of the numbers in Table~\protect \ref{Tab: MtoM0plus}. 
         The shaded area reflects the estimated
         error. For comparison: 
        Standard Monte Carlo SMC
         results~\protect \cite{Teper:1999te} 
        for $\xi=1$ in the middle column and for $\xi=5$ in the right 
      column, where $\xi$ is the anisotropic factor 
        $\xi=\frac{a_s}{a_t}$, with $a_{s(t)}=$ lattice spacing in space 
       (time) direction.}
\label{finalresult}
\end{figure}

Nevertheless we perform a ``final'' estimate
of the glueball spectrum in units of the
$0^+$ glueball by calculating mean values of the numbers of different 
orders and different lattices of Table \ref{Tab: MtoM0plus},
yielding the predictions of Fig.~\ref{finalresult} (left values) .
The errors are deviations of the mean values. We do not take into
account read off errors.

We see that, indeed, we obtain predictions
quite consistent with the Monte Carlo results
of Teper, see Ref.~\cite{Teper:1999te}(middle and right values 
in Fig.~\ref{finalresult}).

\section*{Acknowledgements}

D.S.~acknowledges useful discussions
with Chris Hamer, Gastao Krein, Helmut Kr\"oger and Xiang Luo.

\section*{Appendix}

Within this Appendix we comment on the question whether
it is possible to have a scaling region $\beta \approx 3$
if this region starts only with $\beta \gg 4$ within
the standard lattice Monte Carlo method (see Ref.~\cite{Teper:1999te}).
 
The point is that
a comparison to our Hamiltonian results
involves a rescaling 
of the standard ``Euclidean''  coupling $g_E$
relative to the Hamiltonian coupling $g$.
For $g \rightarrow 0$ this is given by (see Ref.~\cite{Hamer:1996ub})

\begin{eqnarray*}
\beta_{E} &=& \beta + .077 + O(g^2).
\end{eqnarray*}
There is also  rescaling of the Euclidean masses $M_E$ 
relative to  the Hamiltonian masses $M$
(``velocity of light correction'')
\begin{eqnarray*}
M_{E} &=& (1 + 0.084 g^2 +O(g^4)) M.
\end{eqnarray*}

However, one cannot expect that these formulae hold for
$\beta \approx 3$ 
because the difference between $\beta_E$ and $\beta$
becomes very 
large in the strong coupling regime itself.
The strong coupling expansions
are completely different in the 
Euclidean and Hamiltonian formalisms:
the Euclidean expansion has a logarithmic
singularity for $\beta_E = 0$ which
is not present in the Hamiltonian framework.

The rigorous relation between $\beta_E$ and $\beta$
is not known. Consequently for $\beta \approx 3$ a
 direct comparison of masses to Teper's results
appears not to be possible.


\bibliographystyle{plain}

\newpage 

\section*{tables}

\renewcommand{\thefootnote}{\alph{footnote}}

\begin{table}
\caption{Construction of the Loop Space Basis: The functions 
$\Lambda_G^{\delta, k}$ generating $m_{\delta, k}$-dimensional 
subspaces are characterized by simple loop patterns. Altogether 
we have $84$ basis elements in the order $\delta\le 4$ for the 
square lattice.}
\begin{tabular}{lclc}
&&& \\
$\Lambda_G^{\delta, k}$&$m_{\delta, k}$&$\Lambda_G^{\delta, k}$&$m_{\delta, k}$\\
&&& \\
\hline
&&& \\
$\Lambda_G^{0, 1}=1$&1&$\Lambda_G^{4, 1}=$\LOOPEPSBOX{looppicture-geo0-o4-empire8.eps}&1\\
&&$\Lambda_G^{4, 2}=$\LOOPEPSBOX{looppicture-geo0-o4-empire9.eps}&2\\
$\Lambda_G^{1, 1}=$\LOOPEPSBOX{looppicture-geo0-o4-empire1.eps}&1&$\Lambda_G^{4, 3}=$\LOOPEPSBOX{looppicture-geo0-o4-empire10.eps}&7\\
&&$\Lambda_G^{4, 4}=$\LOOPEPSBOX{looppicture-geo0-o4-empire11.eps}&3\\
$\Lambda_G^{2, 1}=$\LOOPEPSBOX{looppicture-geo0-o4-empire2.eps}&1&$\Lambda_G^{4, 5}=$\LOOPEPSBOX{looppicture-geo0-o4-empire12.eps}&5\\
$\Lambda_G^{2, 2}=$\LOOPEPSBOX{looppicture-geo0-o4-empire3.eps}&2&$\Lambda_G^{4, 6}=$\LOOPEPSBOX{looppicture-geo0-o4-empire13.eps}&6\\
&&$\Lambda_G^{4, 7}=$\LOOPEPSBOX{looppicture-geo0-o4-empire14.eps}&4\\
$\Lambda_G^{3, 1}=$\LOOPEPSBOX{looppicture-geo0-o4-empire4.eps}&1&$\Lambda_G^{4, 8}=$\LOOPEPSBOX{looppicture-geo0-o4-empire15.eps}&9\\
$\Lambda_G^{3, 2}=$\LOOPEPSBOX{looppicture-geo0-o4-empire5.eps}&2&$\Lambda_G^{4, 9}=$\LOOPEPSBOX{looppicture-geo0-o4-empire16.eps}&9\\
$\Lambda_G^{3, 3}=$\LOOPEPSBOX{looppicture-geo0-o4-empire6.eps}&4&$\Lambda_G^{4, 10}=$\LOOPEPSBOX{looppicture-geo0-o4-empire17.eps}&10\\
$\Lambda_G^{3, 4}=$\LOOPEPSBOX{looppicture-geo0-o4-empire7.eps}&3&$\Lambda_G^{4, 11}=$\LOOPEPSBOX{looppicture-geo0-o4-empire18.eps}&7\\
&&$\Lambda_G^{4, 12}=$\LOOPEPSBOX{looppicture-geo0-o4-empire19.eps}&6\\
&&&\\
\hline
&&&\\
&&&84\\
&&&
\end{tabular}
\label{Tab: EmpireSquareLattice}
\end{table}

\begin{table}
\caption{Construction of the Loop Space Basis: The functions 
$\Lambda_G^{\delta, k}$ generating $m_{\delta, k}$-dimensional 
subspaces are characterized by simple loop patterns. Altogether 
we have $94$ basis elements in the order $\delta\le 4$ for the 
hexagonal lattice.}
\begin{tabular}{lclc}
&&&\\
$\Lambda_G^{\delta, k}$&$m_{\delta, k}$&$\Lambda_G^{\delta, k}$&$m_{\delta, k}$\\
&&& \\
\hline
&&& \\
$\Lambda_G^{0, 1}=1$&1&$\Lambda_G^{4, 1}=$\LOOPEPSBOX{looppicture-geo1-o4-empire9.eps}&1\\
&&$\Lambda_G^{4, 2}=$\LOOPEPSBOX{looppicture-geo1-o4-empire10.eps}&2\\
$\Lambda_G^{1, 1}=$\LOOPEPSBOX{looppicture-geo1-o4-empire1.eps}&1&$\Lambda_G^{4, 3}=$\LOOPEPSBOX{looppicture-geo1-o4-empire11.eps}&6\\
&&$\Lambda_G^{4, 4}=$\LOOPEPSBOX{looppicture-geo1-o4-empire12.eps}&5\\
$\Lambda_G^{2, 1}=$\LOOPEPSBOX{looppicture-geo1-o4-empire2.eps}&1& $\Lambda_G^{4, 5}=$\LOOPEPSBOX{looppicture-geo1-o4-empire13.eps}&3\\
$\Lambda_G^{2, 2}=$\LOOPEPSBOX{looppicture-geo1-o4-empire3.eps}&2& $\Lambda_G^{4, 6}=$\LOOPEPSBOX{looppicture-geo1-o4-empire14.eps}&5\\
&& $\Lambda_G^{4, 7}=$\LOOPEPSBOX{looppicture-geo1-o4-empire15.eps}&4\\ 
$\Lambda_G^{3, 1}=$\LOOPEPSBOX{looppicture-geo1-o4-empire4.eps}&1&$\Lambda_G^{4, 8}=$\LOOPEPSBOX{looppicture-geo1-o4-empire16.eps}&4\\ 
$\Lambda_G^{3, 2}=$\LOOPEPSBOX{looppicture-geo1-o4-empire5.eps}&2&$\Lambda_G^{4, 9}=$\LOOPEPSBOX{looppicture-geo1-o4-empire17.eps}&6\\
$\Lambda_G^{3, 3}=$\LOOPEPSBOX{looppicture-geo1-o4-empire6.eps}&3&$\Lambda_G^{4, 10}=$\LOOPEPSBOX{looppicture-geo1-o4-empire18.eps}&10\\
$\Lambda_G^{3, 4}=$\LOOPEPSBOX{looppicture-geo1-o4-empire7.eps}&3 & $\Lambda_G^{4, 11}=$\LOOPEPSBOX{looppicture-geo1-o4-empire19.eps}&7 \\
$\Lambda_G^{3, 5}=$\LOOPEPSBOX{looppicture-geo1-o4-empire8.eps}&3&$\Lambda_G^{4, 12}=$\LOOPEPSBOX{looppicture-geo1-o4-empire20.eps}&6\\
&&$\Lambda_G^{4, 13}=$\LOOPEPSBOX{looppicture-geo1-o4-empire21.eps}&8\\
&&$\Lambda_G^{4, 14}=$\LOOPEPSBOX{looppicture-geo1-o4-empire22.eps}&4\\
&&$\Lambda_G^{4, 15}=$\LOOPEPSBOX{looppicture-geo1-o4-empire23.eps}&6\\
&&&\\
\hline
&&&\\
&&&94\\
&&&
\end{tabular}
\label{Tab: EmpireHexagonalLattice}
\end{table}

\begin{table}
\caption{Construction of the Loop Space Basis: The functions 
$\Lambda_G^{\delta, k}$ generating $m_{\delta, k}$-dimensional 
subspaces are characterized by simple loop patterns. Altogether 
we have $58$ basis elements in the order $\delta\le 4$ for the 
triangular lattice.}
\begin{tabular}{lclc}
&&&\\
$\Lambda^{\delta, k}$&$m_{\delta, k}$&$\Lambda^{\delta, k}$&$m_{\delta, k}$\\
&&&\\
\hline
&&&\\
$\Lambda^{0, 1}=1$&1&$\Lambda^{4, 1}=$\LOOPEPSBOX{looppicture-geo2-o4-empire7.eps}&1\\
&&$\Lambda^{4, 2}=$\LOOPEPSBOX{looppicture-geo2-o4-empire8.eps}&2\\
$\Lambda^{1, 1}=$\LOOPEPSBOX{looppicture-geo2-o4-empire1.eps}&1&$\Lambda^{4, 3}=$\LOOPEPSBOX{looppicture-geo2-o4-empire9.eps}&7\\
&&$\Lambda^{4, 4}=$\LOOPEPSBOX{looppicture-geo2-o4-empire10.eps}&3\\
$\Lambda^{2, 1}=$\LOOPEPSBOX{looppicture-geo2-o4-empire2.eps}&1&$\Lambda^{4, 5}=$\LOOPEPSBOX{looppicture-geo2-o4-empire11.eps}&6\\
$\Lambda^{2, 2}=$\LOOPEPSBOX{looppicture-geo2-o4-empire3.eps}&2&$\Lambda^{4, 6}=$\LOOPEPSBOX{looppicture-geo2-o4-empire12.eps}&9\\
&&$\Lambda^{4, 7}=$\LOOPEPSBOX{looppicture-geo2-o4-empire13.eps}&10\\
$\Lambda^{3, 1}=$\LOOPEPSBOX{looppicture-geo2-o4-empire4.eps}&1&$\Lambda^{4, 8}=$\LOOPEPSBOX{looppicture-geo2-o4-empire14.eps}&8\\
$\Lambda^{3, 2}=$\LOOPEPSBOX{looppicture-geo2-o4-empire5.eps}&2&&\\
$\Lambda^{3, 3}=$\LOOPEPSBOX{looppicture-geo2-o4-empire6.eps}&4&&\\
&&&\\
&&&\\
\hline
&&&\\
&&&58\\
&&&
\end{tabular}
\label{Tab: EmpireTriangularLattice}
\end{table}

\begin{table}
\caption{Lattice Angular Momenta: The number of basis elements 
${\chi^\alpha}$ for the different spins on the square lattice.}
\begin{tabular}{ccccccccc}
&&&&&&&&\\
order&$\delta\le$0&$\delta\le$1&$\delta\le$2&$\delta\le$3&$\delta\le$4&$\delta\le$5&$\delta\le$6&$\delta\le$7\\
&&&&&&&&\\
\hline
&&&&&&&&\\
$\chi^{\alpha}$&$1$&$2$&$5$&$15$&$84$&$557$&$4942$&$47751$\\
&&&&&&&&\\
\hline
&&&&&&&&\\
$\Pi^{0^+}\chi$&$1$&$2$&$5$&$15$&$84$&$557$&$4942$&$47751$\\
$\Pi^{0^-}\chi$&$0$&$0$&$0$&$1$&$31$&$377$&$4220$&$45132$\\
$\Pi^{2^+}\chi$&$0$&$0$&$2$&$8$&$65$&$497$&$4759$&$47073$\\
$\Pi^{2^-}\chi$&$0$&$0$&$0$&$4$&$42$&$425$&$4386$&$45790$\\
$\Pi^{1^{\pm}}\chi$&$0$&$0$&$0$&$8$&$79$&$874$&$8841$&$92340$\\
&&&&&&&&
\end{tabular}
\label{Tab: PiChiSquareLattice}
\end{table}

\begin{table}
\caption{Lattice Angular Momenta: The number of basis elements 
${\chi^\alpha}$ for the different spins on the hexagonal lattice.}
\begin{tabular}{ccccccccc}
&&&&&&&&\\
order&$\delta\le$0&$\delta\le$1&$\delta\le$2&$\delta\le$3&$\delta\le$4&$\delta\le$5&$\delta\le$6&$\delta\le$7\\
&&&&&&&&\\
\hline
&&&&&&&&\\
$\chi$&$1$&$2$&$5$&$17$&$94$&$677$&$6430$&$67036$\\
&&&&&&&&\\
\hline
&&&&&&&&\\
$\Pi^{0^+}\chi$&$1$&$2$&$5$&$17$&$94$&$677$&$6430$&$67036$\\
$\Pi^{0^-}\chi$&$0$&$0$&$0$&$1$&$35$&$464$&$5613$&$63947$\\
$\Pi^{3^+}\chi$&$0$&$0$&$0$&$4$&$43$&$512$&$5728$&$64573$\\
$\Pi^{3^-}\chi$&$0$&$0$&$0$&$6$&$52$&$573$&$5989$&$65854$\\
$\Pi^{1^{\pm}}\chi$&$0$&$0$&$0$&$8$&$91$&$1078$&$11691$&$130381$\\
$\Pi^{2^{\pm}}\chi$&$0$&$0$&$2$&$12$&$120$&$1128$&$12008$&$130924$\\
&&&&&&&&
\end{tabular}
\label{Tab: PiChiHexagonalLattice}
\end{table}

\begin{table}
\caption{Lattice Angular Momenta: The number of basis elements 
${\chi^\alpha}$ for the different spins on the triangular lattice.}
\begin{tabular}{ccccccccc}
&&&&&&&&\\
order&$\delta\le$0&$\delta\le$1&$\delta\le$2&$\delta\le$3&$\delta\le$4&$\delta\le$5&$\delta\le$6&$\delta\le$7\\
&&&&&&&&\\
\hline
&&&&&&&&\\
$\chi$&$1$&$2$&$5$&$12$&$58$&$341$&$2958$&$27885$\\
&&&&&&&&\\
\hline
&&&&&&&&\\
$\Pi^{0^+}\chi$&$1$&$2$&$5$&$12$&$58$&$341$&$2958$&$27885$\\
$\Pi^{0^-}\chi$&$0$&$0$&$0$&$1$&$21$&$249$&$2601$&$26857$\\
$\Pi^{3^+}\chi$&$0$&$0$&$0$&$1$&$22$&$250$&$2603$&$26859$\\
$\Pi^{3^-}\chi$&$0$&$1$&$2$&$9$&$41$&$324$&$2807$&$27734$\\
$\Pi^{1^{\pm}}\chi$&$0$&$0$&$0$&$7$&$56$&$562$&$5393$&$54550$\\
$\Pi^{2^{\pm}}\chi$&$0$&$0$&$2$&$9$&$71$&$577$&$5538$&$54695$\\
&&&&&&&&
\end{tabular}
\label{Tab: PiChiTriangularLattice}
\end{table}

\clearpage

\begin{table}
\caption{The mass ratios of various glueball states $J^P$ relative to the
$0^+$ state for the square, the square improved, the hexagonal and the triangular lattice 
in 7th and 8th order in the scaling region.} 
\begin{tabular}{ccccccccc}
&&&&&&&&\\
&\multicolumn{2}{c}{square}&\multicolumn{2}{c}{square impr.}&\multicolumn{2}{c}{hexagonal}&\multicolumn{2}{c}{triangular}\\
$J^P$&\multicolumn{2}{c}{$\beta=3$}&\multicolumn{2}{c}{$\beta=1.7$}&\multicolumn{2}{c}{$\beta=3$}&\multicolumn{2}{c}{$\beta=3$}\\
&$\delta=7$&$\delta=8$&$\delta=7$&$\delta=8$&$\delta=7$&$\delta=8$&$\delta=7$&$\delta=8$\\
&&&&&&&&\\
\hline
&&&&&&&&\\
$0^{+}$&$1$&$1$&$1$&$1$&$1$&$1$&$1$&$1$\\
$0^{+*}$&$-$&$1.47$&$1.43$&$1.41$&$1.51$&$1.42$&$-$&$1.76$\\
&&&&&&&&\\
$0^{-}$&$1.78\footnotemark[1]$&$2.00\footnotemark[1]$&$1.82\footnotemark[1]$&$2.01\footnotemark[1]$&$1.65\footnotemark[1]$&$2.00\footnotemark[1]$&$2.13\footnotemark[1]$&$2.26\footnotemark[1]$\\
$0^{-*}$&$2.01\footnotemark[1]$&$2.06\footnotemark[1]$&$2.06\footnotemark[1]$&$2.07\footnotemark[1]$&$2.08\footnotemark[1]$&$2.15\footnotemark[1]$&$2.18\footnotemark[1]$&$2.32\footnotemark[1]$\\
$0^{-**}$&$2.30\footnotemark[1]$&$2.20\footnotemark[1]$&$2.34\footnotemark[1]$&$2.23\footnotemark[1]$&$2.25\footnotemark[1]$&$2.20\footnotemark[1]$&$2.53\footnotemark[1]$&$2.53\footnotemark[1]$\\
&&&&&&&&\\
$1^{\pm}$&$2.11$&$2.10$&$2.14$&$2.10$&$2.17$&$2.15$&$2.07\footnotemark[2]$&$2.04\footnotemark[2]$\\
$1^{\pm *}$&$2.23$&$2.22\footnotemark[1]$&$2.27$&$2.25$&$2.26$&$2.17\footnotemark[1]$&$2.29\footnotemark[2]$&$2.12\footnotemark[2]$\\
$1^{\pm **}$&$2.27$&$2.23$&$2.30$&$2.26$&$2.26$&$2.19\footnotemark[1]$&$2.39$&$2.35$\\
&&&&&&&&\\
$2^+$&$1.56$&$1.61$&$1.57$&$1.60$&$$&$$&$$&$$\\
$2^{+*}$&$1.77$&$1.87$&$1.78$&$1.83$&$$&$$&$$&$$\\
$2^{+**}$&$1.78$&$-$&$1.82$&$-$&$$&$$&$$&$$\\
&&&&&&&&\\
$2^-$&$1.68$&$1.66$&$1.69$&$1.66$&$$&$$&$$&$$\\
$2^{-*}$&$1.77$&$-$&$1.81$&$-$&$$&$$&$$&$$\\
$2^{-**}$&$2.11$&$-$&$2.08$&$-$&$$&$$&$$&$$\\
&&&&&&&&\\
$2^{\pm}$&$$&$$&$$&$$&$1.60$&$1.62$&$1.66$&$1.65$\\
$2^{\pm *}$&$$&$$&$$&$$&$1.65$&$1.83$&$1.97$&$1.99$\\
$2^{\pm **}$&$$&$$&$$&$$&$1.65$&$1.96$&$2.03$&$2.05$\\
&&&&&&&&\\
$3^{+}$&$$&$$&$$&$$&$2.05$&$2.05$&$2.20$&$2.27$\\
$3^{+*}$&$$&$$&$$&$$&$2.26$&$2.16\footnotemark[1]$&$2.55$&$2.52$\\
$3^{+**}$&$$&$$&$$&$$&$2.31$&$2.18\footnotemark[1]$&$2.63$&$2.61$\\
&&&&&&&&\\
$3^{-}$&$$&$$&$$&$$&$2.06$&$2.08$&$1.69\footnotemark[2]$&$1.69\footnotemark[2]$\\
$3^{-*}$&$$&$$&$$&$$&$2.27$&$2.52\footnotemark[1]$&$2.09\footnotemark[2]$&$1.97\footnotemark[2]$\\
$3^{-**}$&$$&$$&$$&$$&$2.29$&$2.19\footnotemark[1]$&$2.23$&$2.21$\\
&&&&&&&&
\end{tabular}
\footnotetext[1]{predominantly $2$-cluster states}
\footnotetext[2]{level crossing}
\label{Tab: MtoM0plus}
\end{table}

\clearpage

\section*{figures}

\end{document}